\documentclass[preprints,article,accept,moreauthors,pdftex]{Definitions/mdpi} 
\firstpage{1} 
\pubvolume{1}
\issuenum{1}
\articlenumber{0}
\pubyear{2022}
\copyrightyear{2022}
\datereceived{} 
\dateaccepted{} 
\datepublished{} 
\hreflink{https://doi.org/} % If needed use \linebreak
\pdfoutput=1

\usepackage[version=4]{mhchem}
\Title{Ammonia and Phosphine in the Clouds of Venus as Potentially Biological Anomalies}

\TitleCitation{Ammonia and Phosphine in the Clouds of Venus as Potentially Biological Anomalies}

\Author{Carol E. Cleland $^{1,*,\ddagger}$\orcidB{}, Paul B. Rimmer $^{2,\dagger,\ddagger}$\orcidA{}}

\AuthorNames{P.B. Rimmer, C.E. Cleland}

\AuthorCitation{Cleland, C.~E., Rimmer, P.~B.; }
\address{%

$^{1}$ \quad Center for the Study of Origins, Department of Philosophy, University of Colorado, Boulder, CO 80309, USA; carol.cleland@colorado.edu\\
$^{2}$ \quad Cavendish Laboratory, University of Cambridge, JJ Thomson Avenue, Cambridge CB3 0HE, UK; pbr27@cam.ac.uk}

\corres{Correspondence: carol.cleland@colorado.edu; Tel.:  +1-303-735-4558 (C.E.C.)}

\firstnote{Correspondence: pbr27@cam.ac.uk; Tel.: +44-01223-337-308 (P.B.R.)} 
\secondnote{These authors contributed equally to this work.}
\abstract{We are of the opinion that several anomalies in the atmosphere of Venus provide evidence of yet-unknown processes and systems that are out of equilibrium. The investigation of these anomalies on Venus should be open to the wide range of explanations, including unknown biological activity. We provide an overview of two anomalies, the tentative detection of ammonia and phosphine in Venus's atmosphere. These anomalies fly in the face of the tacit assumption that the atmosphere of Venus must be in chemical redox equilibrium, an assumption connected to the belief that Venus is lifeless. We then discuss several major past discoveries in astronomy, biology and geology, which lead to the abandonment of certain assumptions held by many scientists as though they were well-established principles. The anomalies of ammonia and phosphine in the atmosphere of Venus are placed in the context of these historical discoveries. This context supports our opinion that persistence by the community in the exploration of these anomalies with a skeptical eye towards tacit assumptions will increase the chances of making profound discoveries about the atmosphere of Venus and the diverse and often strange nature of planetary environments.}


\begin{document}

%%%%%%%%%%%%%%%%%%%%%%%%%%%%%%%%%%%%%%%%%%

\section{Introduction}
\label{sec:intro}

This paper discusses two of several anomalies in the atmosphere of Venus and shows that the scientific community responds to these anomalies in the same manner that it has responded to anomalous phenomenon in the past--phenomena that were unrecognized for what they represented and foreshadowed important scientific discoveries. We opine that, if the past is any indication, investigating these Venusian anomalies will lead to important new discoveries about Venus, planetary environments, and perhaps even novel possibilities for extraterrestrial life. These investigations will be facilitated by the scientific community re-examining certain strongly held but non-fundamental assumptions about Venus’ atmosphere. These assumptions, functioning for many scientists as well-established principles, influence the ways in which observations and experimental results concerning Venus are interpreted, blinkering researchers to possibilities that they might otherwise consider.

\section{Anomalies in the clouds of Venus: Ammonia and Phosphine}
\label{sec:venus-anomalies}

The Venera, Pioneer and VeGa probes all discovered various phenomena that are difficult to reconcile with each other, and impossible to reconcile with the widely accepted assumption (amounting to a principle) among some planetary scientists that the atmosphere of Venus is in chemical redox equilibrium. By ``redox disequilibrium'', we mean a separation between oxidizing and reducing species, as has been proposed by Hitchcock \& Lovelock to be a means by which to detect life \cite{Hitchcock1967}. To better describe redox disequilibrium, we follow the convention of Catling \& Kasting \cite{Catling2017}. We quantify redox with reference to \ce{H2} and standard ``redox neutral'' species as follows: if \ce{H2} is assigned a redox value of $-1$ and \ce{H2O} a redox value of $+0$, then \ce{O2} must have a value of $+2$ when considering the stoichiometry of 2 \ce{H2} + \ce{O2} $\rightarrow$ 2 \ce{H2O}. Likewise, if \ce{CO2} is assigned the value of $+0$, and since \ce{CO} + \ce{H2O} $\rightarrow$ \ce{CO2} + \ce{H2}, \ce{CO} has a redox value of $-1$. If \ce{SO2} is given a redox value of $+0$, \ce{H2S} has a redox value of $-3$ and \ce{H2SO4} a value of $+1$. If \ce{N2} is assigned the value $+0$, \ce{NH3} must have the redox value of $+3/2$. Redox disequilibrium is when there are significant abundances of two species, one of which carries a negative value, and the other a positive value, together in the atmosphere. Significant amounts of \ce{NH3} with \ce{H2SO4} qualifies as significant redox disequilibrium.

Redox disequilibrium is not an all-or-nothing state of affairs. It comes in degrees. It is debatable what degree of redox disequilibrium qualifies as ``significant disequilibrium''. Some attempts have been made to discover this boundary, when considering life-detection in the context of thermodynamic-chemical disequilibrium \cite{Krissansen2016,Schwieterman2018}. These attempts illustrate the need for a detailed and holistic view of the planetary environment, and will be difficult to establish based on a sample size of one. Because of the proposal by Hitchcock \& Lovelock, claims of redox disequilibrium has been inextricably linked with claims of alien life, and this can affect how redox disequilibrium is treated as an anomaly.

Anomalous phenomena observed in the atmosphere of Venus include the depletion of sulfur and water in the clouds of Venus \cite{Yung1982}, the strange behavior of \ce{SO2} near the surface \cite{Bertaux1996}, and \ce{H2O} in the clouds \cite{Ignatiev1999,DeBergh2006}, the detection of \ce{O2}, \ce{H2S} and \ce{CH4} in the clouds and non-detection of \ce{O2} above the clouds \cite{Oyama1980,Donahue1993,Krasnopolsky2006}, the depletion of \ce{OCS} below the clouds \cite{Yung2009,Bierson2020}, the detection of phosphorous clouds below sulfur clouds \cite{Andreychikov1987,Krasnopolsky1989,Krasnopolsky2006}, and a host of phenomena in the surface mineralogy: the mineral composition does not appear to be in equilibrium with the lower atmosphere \cite{Rimmer2021}. For an overview and discussion of these ostensibly incompatible chemicals in the clouds of Venus, see Petkowski et al. \cite{Petkowski2022}. Recent efforts from the Venus community to seriously account for these anomalies and explore both biotic and abiotic explanations comprises an entire collection in {\it Astrobiology}: the Venus Collection. See Limaye et al. \cite{Limaye2021} for an introduction to this collection.

The anomaly of \ce{O2}, \ce{H2S} and \ce{CH4} together with \ce{H2SO4}, or \ce{PH3} or \ce{NH3} together with \ce{H2SO4}, in the clouds of Venus, qualifies as significant redox disequilibrium, an anomaly in the clouds of Venus that has been considered as evidence of biological activity. Two anomalies, the ammonia (\ce{NH3}) and phosphine (\ce{PH3}) detected in the atmosphere of Venus, are the focus of this discussion.  They provide revealing illustrations of the way in which scientific communities typically respond to the detection of anomalies. We discuss these anomalies in more detail below.

\subsection{Ammonia}
\label{sec:ammonia}

The presence of ammonia (\ce{NH3}) in the atmosphere of Venus, between 30 and 45 km atmospheric height, was inferred from the color change of bromophenol blue in calcined silica gel observed on Venera 8 in 1972 at concentrations of 100-1000 ppm \cite{Surkov1973,Surkov1977}. These observations were dismissed by Goettel \& Lewis in 1974 primarily on the basis that ``the high \ce{NH3} mixing ratios reported by the Soviet Venera 8 landing probe appear to be inconsistent with the observed abundances of other gases in the Venus atmosphere.'' \cite{Goettel1974}. In other words, the detection of \ce{NH3} was effectively dismissed because it was anomalous in the sense that it was inconsistent with other empirical findings about the atmosphere of Venus. Namely, they were inconsistent with the set of findings that supported the assumption that Venus’ atmosphere is in chemical equilibrium.  This was effectively the only reason given at the time for dismissing the Venera 8 results, in spite of empirically-based concerns, e.g. about the potential for interaction of $300^\circ$C concentrated sulfuric acid with bromophenol blue to produce a false-positive detection of \ce{NH3} \cite{Petkowski2022}. 

In 1978, the Pioneer Venus multiprobe (consisting of one large probe and three smaller probes) traveled through the atmosphere of Venus. Readings from the large probe’s neutral mass spectrometer (LNMS) were consistent with the presence of \ce{NH3}, but, possibly because the presence of \ce{NH3} was earlier dismissed and therefore unexpected, these mass spectrometry readings were not noticed or ignored (in any case, they went unreported) until reanalysis decades later by Mogul et al. \cite{Mogul2021}. There is, however, a hypothesis that can explain the baffling presence of \ce{NH3} and other anomalies in the clouds of Venus, such as the other reduced species and in-cloud \ce{O2} \cite{Bains2021}.  This hypothesis involves attributing the detected, atmospheric, chemical redox disequilibrium to life, as opposed to insisting that it must be explained in terms of poorly understood or unknown abiotic chemical reactions.

It is important to note that the scientific community has not been uniform in the dismissal of redox disequilibrium or consideration of the life hypothesis. Several researchers have seriously investigated the life hypothesis and other hypotheses to explain the chemistry in the clouds \cite{Grinspoon1997,Schulze-Makuch2002,Schulze-Makuch2004,Grinspoon2007,Limaye2021}. This work spans the time between the first claimed detection of \ce{NH3} and the recent reanalysis of the Pioneer data. In particular, Schulze-Makuch et al. \cite{Schulze-Makuch2004} highlight the claimed redox disequilibrium: ``the atmosphere is in chemical disequilibrium, with \ce{H2} and \ce{O2}, and \ce{H2S} and \ce{SO2} coexisting.''

\subsection{Phosphine}
\label{sec:phosphine}

A claimed detection of an absorption line at 267 GHz with the Atacama Large Millimeter/sub-millimeter Array (ALMA) and the James Clerk Maxwell Telescope (JCMT) was attributed to \ce{PH3} in the clouds of Venus by Greaves et al. \cite{Greaves2021}. This is an especially provocative anomaly. Prior to the reported detection of PH3 on Venus, phosphine had been proposed as a potentially definitive atmospheric biosignature on rocky planets \cite{Sousa2020}. No known abiotic source can explain its presence in the clouds of Venus \cite{Bains2021b}. 

Naturally, and rightly, the veracity of the detection of phosphine was challenged: 

First, the detection itself was challenged. The ALMA detection has not held up well \cite{Snellen2020,Akins2021,Villanueva2021}, though see also Greaves et al. \cite{Greaves2021b,Greaves2021c}, but the JCMT detection has been independently confirmed multiple times \cite{Lincowski2021,Villanueva2021}, though see also \cite{Thompson2021}. The detection of phosphine was then attributed to a less anomalous molecule, sulfur dioxide (\ce{SO2}) \cite{Lincowski2021}, already known and expected to be in the atmosphere of Venus. The amount of \ce{SO2} required to explain the signal, however, appears to be inconsistent with contemporaneous independent measurements of \ce{SO2} \cite{Greaves2021b,Greaves2022}. There is also some independent evidence of \ce{PH3} in the clouds from recent reanalysis of Pioneer Venus LNMS data \cite{Mogul2021}. Other features of \ce{PH3}, however, have not been detected \cite{Cordiner2022}.

There have also been objections to the inferred location of the \ce{PH3} by Greaves et al. Lincwoski et al. \cite{Lincowski2021} claim the \ce{PH3} must be above the clouds to produce the 267 GHz absorption, and independent reanalysis has confirmed this claim \cite{Villanueva2021,Greaves2022}. Stipulating that the 1 mm feature is explained by \ce{PH3}, the location of \ce{PH3} above the clouds makes its presence even more anomalous, because \ce{PH3} is even less stable outside the clouds, where destructive ultraviolet light is much more intense \cite{Bains2021b}. In short, if \ce{PH3} is above the clouds, something must be constantly replenishing the \ce{PH3}.  There is no known abiotic chemical mechanism that can replenish the \ce{PH3} fast enough. Terrestrial biotic fluxes would be sufficient to replenish the \ce{PH3} in the clouds, but not above the clouds. For this reason, it also makes the claim of life as we know it far less plausible, because even known biology would not be able to produce enough \ce{PH3} to explain these observations. It is also challenging to reconcile with independent upper limits for the above-cloud \ce{PH3} \cite{Encrenaz2020,Trompet2021}. 

Finally, there have been some attempts to explain the presence of \ce{PH3} on Venus through known abiotic chemistry \cite{Omran2021,Truong2021}. So far, however, none of these attempts have been able to satisfactorily explain the presence of \ce{PH3}, either because they do not predict enough \ce{PH3} to explain the observations, especially above the clouds, or because they predict other molecules, especially water vapor, in large amounts that contradict observations \cite{Bains2021b,Bains2022,Bains2022b}. Life using expected energy-metabolic pathways produce similar contradictions \cite{Jordan2022}. If there is observable \ce{PH3} above the clouds of Venus, we are left with two explanations: life-as-we-don’t-know-it and nonlife-as-we-don’t-know-it. Both of these possibilities should be seriously considered, and neither should be clearly favored over the other until there’s more data. Instead, more predictions should be made, and more data should be gathered, especially {\it in situ} data from upcoming missions (e.g. \cite{Garvin2022,Seager2022,French2022}). 

Yet the latter (abiotic) hypothesis remains the preferred option among planetary scientists. As discussed below, this is typical of scientific approaches to explaining anomalous findings, that is, findings that challenge widely accepted scientific assumptions, such as that Venus’ atmosphere should be in chemical equilibrium, which in turn rests upon the assumption that there is no life on Venus.  For as Sara Seager \cite{Seager2014}, among others, argues, the maintenance of a planetary atmosphere in a state of chemical disequilibrium is a promising sign of life. The main difference is that she is talking about searching for evidence of biology on extrasolar planets.  Few astrobiologists anticipated finding such evidence on Earth’s next-door neighbor, Venus. 

\section{The Role of Anomalies in Scientific Discovery}
\label{sec:role}

A (scientific) anomaly is a phenomenon that is surprising (unanticipated or unexpected) in the context of a widely accepted theoretical framework—a collection of principles, some grounded in fundamental theory and others based on highly plausible, ancillary doctrines—for understanding phenomena of the sort concerned. 

Early to mid-twentieth century philosophers of science rarely spoke of ``anomalies'', treating them as an especially difficult to explain (and hence very surprising) species of failed prediction; the difference between an anomaly and a failed prediction was thus viewed as primarily psychological or sociological, as opposed to logical. Thomas Kuhn \cite{Kuhn1962} turned the tables, placing the concept of anomaly at the center of a radically new account of scientific practice. According to Kuhn, anomalies are initially viewed as puzzles that can be explained (given sufficient ingenuity) using the resources provided by the pertinent dominant ``paradigm'' (Kuhn’s concept of a theoretical framework).  If the puzzling phenomenon persists in thwarting efforts to explain it in this way, the scientific community may (there is no guarantee) enter a Kuhnian ``crisis''. Kuhn is unclear as to whether a puzzling phenomenon can be recognized as “anomalous” during a crisis, that is, in the absence of paradigm change. He sometimes suggests that a phenomenon cannot be recognized as anomalous until the paradigm is replaced in a scientific revolution by a new paradigm and other times suggests that the transition to a crisis (weakening the grip of a paradigm on the scientific mind) results in a puzzle being recognized as anomalous \cite{Kuhn1962}. On Kuhn’s account, commitment to a paradigm makes it difficult if not impossible to recognize the anomalous nature of a phenomenon. 

Kuhn’s account of science is no longer widely accepted among philosophers of science. It faces many problems, a discussion of which is beyond the scope of this paper.  Nonetheless, Kuhn deserves credit for recognizing that anomalies amount to more than failed predictions and, most importantly, play central roles in the process of scientific discovery. As any scientist recognizes failed predictions are not surprising.  They happen frequently and are typically explained in terms of a lack of pertinent controls (lab experiments) or poorly understood conditions causally influencing the phenomenon being observed (field studies, which include astronomy and planetary science).  While a failed prediction may eventually be recognized as anomalous, many anomalies do not represent failed predictions. Anomalies are surprising because they fall outside the scope of expectations (articulated in prediction and explanation) induced by widely accepted theoretical frameworks (see \cite{Stanford2006}). The following illustration from the history of astronomy highlights this important point.

\subsection{A Case Study: The discovery of Uranus the planet}
\label{sec:uranus}

Until 1781, some of the best astronomers in Europe (including Britain’s first Astronomer Royal, John Flamsteed, and French astronomer Pierre Lemonnier) had catalogued the celestial object occupying a location that we now know was occupied by the planet Uranus as a star; see Evans \cite{Evans2021} for more historical detail.  Using a new, more powerful, telescope of his own devising, William Herschel noticed that the object was anomalous for a star:  It was disk shaped and moved among (as opposed to with) the stars.  Herschel announced his discovery, musing that it was probably a comet while noting that it lacked a cometary gaseous tail. Astronomers throughout Europe trained the same telescopes with which they’d previously reported observing a star and announced that Herschel was correct:  The object was anomalous for a star.  The point is their ``inferior'' telescopes didn’t prevent them from recognizing earlier that Uranus lacked the diagnostic characteristics of a star.  Why, then, did they report a star in this location? The most likely explanation is that they believed the then widely accepted principle that there are no planets beyond Saturn’s orbit. 

This case illustrates two important points about anomalies and their role in scientific discovery. First, the discovery that Uranus isn't a star didn't result from a failed prediction nor did it contradict fundamental theory. In hindsight, one could reconstruct Herschel’s discovery as a failed prediction, based on the then diagnostic characteristics of a star, but this isn't how it happened.  Herschel wasn't testing the conjecture that Uranus is a star. While scanning the night sky with a new telescope, he literally stumbled onto an object that he couldn't explain within the then dominant scientific framework for understanding celestial phenomena:  It lacked the diagnostic characteristics of a star, and moreover its correct interpretation challenged a widely accepted principle (accepted by Herschel) that there are no planets beyond Saturn. Significantly, the object wasn’t anomalous vis-\'{a}-vis fundamental theory, namely, Newton’s theory of universal gravitation. 

Second, ancillary principles can thwart the recognition of anomalies just as effectively as fundamental principles. Kuhn \cite{Kuhn1962}, who focused his account on major theory change (scientific revolutions), attributes the principle that there are no planets beyond Saturn to lingering  influences of the Ptolemaic, Earth centered, model of the universe, which seems implausible since the Copernican model was widely accepted by this time.  Moreover, Herschel, like most astronomers in the late eighteenth century, was a Copernican. Yet he didn't take the conjecture that Uranus is a planet as seriously as the conjecture that it is a comet, despite noting the lack of a gaseous tail.  This suggests that the assumption that there are no planets beyond Saturn was an ancillary, as opposed to fundamental, principle about the cosmos.  

After Anders Lexell computed the orbit of Uranus, around 6 months later, and announced that it was planetary (vs. cometary), astronomers throughout Europe, using the same telescopes with which they had originally identified Uranus as a star, began discovering other objects having planetary orbits. These ``planetoids'' (large asteroids) were visible through their “inferior” telescopes but were previously unrecognized for what they represent because, until the discovery of Uranus the planet, astronomers weren't expecting to find additional objects having planetary orbits. 

\subsection{How ancillary (vs. fundamental) principles can blind scientists to alternative explanations of anomalies}
\label{sec:anc}

The history of science is littered with examples of anomalies that played key roles in important scientific discoveries but were unrecognized for what they represent because they violated widely accepted ancillary principles. Some additional illustrations are: In biochemistry, the discovery of ribozymes, which violated a widely accepted version of the ``central dogma of molecular biology'', which held that only proteins (not nucleic acids, like RNA) could be biologically effective catalysts \cite{Nobel1989}; In geology, the inexplicable similarity in the shape of Africa and South America coupled with mysterious geological and paleontological similarities on the West coast of Europe and Africa and the east coast of the Americas, which couldn't be explained by the widely held dogma that continental landmasses don’t move \cite{Oreskes1999}; In pre-Darwinian biology, the classification of whales as fish on the basis of their external morphology, despite the fact that their internal structure more closely resembled mammals \cite{Burnett2007}. 

None of these cases challenged fundamental theory. They violated widely accepted, highly plausible, ancillary principles. Rejecting the central dogma of molecular biology didn't violate the basic laws of chemistry. Insofar as only proteins had thus far exhibited catalytic properties it was highly plausible. The puzzle over the shapes of Africa and South America challenged the highly plausible assumption that continental landmasses can’t move:  How could such enormous things do so?  Lastly, the reclassification of whales as mammals occurred before the acceptance of Darwin’s theory of evolution.  Carl Linnaeus, the father of modern taxonomy, followed his predecessors in classifying them as fish on the basis of “common sense” (diagnostic criteria based on external morphology and lifestyle \cite{Romero2012}).

\section{Ammonia and Phosphine as Potentially Biological Anomalies}
\label{sec:nh3-ph3-anom}

As discussed in Section \ref{sec:venus-anomalies}, detecting ammonia and phosphine in the atmosphere of Venus challenges a widely accepted assumption among planetary scientists, namely, that Venus' atmosphere should be in chemical redox equilibrium. In the case of ammonia, commitment to this ancillary principle might have distracted from the fundamental behavior of a specific detector's chemistry with concentrated sulfuric acid in the clouds of Venus. The assumption that Venus' atmosphere should be in equilibrium is based on a widely shared belief, functioning in debates over explaining the tentative detections of ammonia and phosphine as an ancillary principle, namely, that life doesn't exist on Venus; the conjecture that there is life on Venus doesn't violate basic principles of planetary science. The issue is not whether a life explanation for the anomalous ammonia and phosphine is correct but whether it is being ignored as a result of tacit commitment to a poorly grounded ancillary principle. 

It is useful to consider another example of a claimed detection of an anomalous gas-phase species in a planetary atmosphere. Methane has been claimed to be present on Mars in part-per-billion (ppb) amounts that vary both temporally and spatially \cite{Mumma2009}, and those claims have been challenged \cite{Zahnle2011}, including by one of the authors of this opinion \cite{Gillen2020}. Not just the presence of ppb levels of levels of methane but its variation over such short timescales is highly anomalous, requiring first more robust evidence to establish the presence of ppb levels of methane and its rapid variation. If ppb levels of methane on Mars are confirmed, the anomaly should be addressed with an open mind to both biotic and abiotic explanations. The greater anomaly concerning methane on Mars may well be the rapid destruction of methane, rather than its mere presence in trace quantities \cite{Zahnle2011}.

Cleland \cite{Cleland2019,Cleland2019b} argues that a scientifically more promising strategy for searching for extraterrestrial life is to search for potentially biological anomalies, a strategy anticipated by David Grinspoon \cite{Grinspoon2003}:

\begin{quote}
We can best explore astrobiologically by roaming widely and keeping a sharp eye out for anomalous order of any kind. This could include strange, nonequilibrium mixtures of gases (or, conversely, too much equilibrium in places where other known processes are creating disequilibrium!), strange mechanical shapes and assemblages, or rhythmic environmental changes without any obvious cause. Such anomalous order will indicate either an interesting nonbiological process that we need to learn about† or that we have at last found new life.
\end{quote}

Contemporary approaches for searching for extraterrestrial life focus on using defining criteria or ``agnostic biosignatures''. The problem with both strategies (which are closely related) is that familiar Earth life represents a single example that may be unrepresentative of life elsewhere in the universe. 

We just don't know which characteristics of familiar Earth life can be safely generalized to all life in the universe in a biologically informative way.  Scientists are currently in no position to discriminate which of the universal characteristics of familiar Earth life are common (let alone fundamental) to life considered generally.  There is also the problem of the ``Goldilocks level of abstraction ''(\cite{Cleland2019b}, Ch. 4.3).  For example, while it is highly plausible that all life engages in metabolism it is difficult to provide a characterization of metabolism that rules out abiotic system, such as hurricanes, which sustain themselves by extracting energy and matter from the environment, while at the same time avoiding being too Earth-centric, that is, specific to the biochemistry of familiar life.  Popular ``agnostic biosignatures'' are supposed to avoid the well-known problems with defining life, but to the extent that they at least tacitly (sometimes even explicitly) specify necessary conditions for life, they are not so much ``agnostic'' as based on incomplete or partial definitions of life; the term ``agnostic'' is a misnomer, misleading readers into thinking that they are more neutral (vis-\'{a}-vis using familiar Earth life as a basis for searching for extraterrestrial life) than they are.

Cleland’s (\cite{Cleland2019}; \cite{Cleland2019b}, Ch. 8) proposal that astrobiologists search for potentially biological anomalies is truly agnostic, avoiding the problems of using Earth-centric defining criteria or ``agnostic'' biosignatures in a search for life that may differ from familiar life in poorly understood or even as yet unconceived ways (see also \cite{Stanford2006}). Potentially biological anomalies may resemble familiar life in provocative ways not previously entertained as diagnostic of life and they may also differ from familiar life in ways suggestive of abiotic processes; they may defy classification as life or nonlife. Once identified as a potentially biological anomaly, such a system becomes a focus of further research to discriminate whether it represents an unfamiliar form of biology or an unfamiliar abiotic processes. 

An additional advantage of this approach is that it holds forth the promise of expanding the available biosignatures for searching for extraterrestrial life. Potentially biological anomalies that persist in resisting classification as the outcome of abiotic processes, despite not representing clear cut cases of life, provide tools for exploring novel possibilities of alien life on other worlds. The persistence of these anomalies can provide evidence for the presence of life, in a Bayesian sense of rendering the life hypothesis (even slightly) more likely after considering this new information. Persistent and potentially biological anomalies can therefore be considered as (tentative) biosignatures within a Bayesian framework (a framework as constructed by Catling et al. \cite{Catling2018} and Walker et al. \cite{Walker2018}).

If present, the anomalous ammonia and phosphine in the atmosphere of Venus represent a case of a potentially biological anomaly.  For as Section 2.2 discusses, they are difficult to classify as either the product of life or abiotic processes. If phosphine in Venus' atmosphere is being caused by life, then Venusian life differs in significant ways from familiar Earth life. On the other hand, one cannot rule out the possibility that the phosphine anomaly is the product of poorly understood or unknown abiotic processes in the highly complex atmosphere of Venus. In short, we argue that neither hypothesis should be favored over the other, because it is difficult to reliably assign likelihoods to unknown mechanisms on a solely empirical basis. Extra-empirical considerations, such as parsimony, could lead one to rationally favor abiotic explanations over biotic explanations.  Both possibilities should be seriously considered and carefully investigated. If it turns out that there is significant redox disequilibrium in and above the clouds of Venus, and that this disequilibrium has an abiotic explanation, then this will effectively falsify Hitchcock \& Lovelock's hypothesis.

In sum, the (still tentative) detections of anomalous amounts of ammonia and phosphine in Venus’ atmosphere hold forth the promise of expanding our current scientific understanding of how (both) biological and abiotic processes affect planetary atmospheres.  For even if it turns out that these chemical anomalies can be explained abiotically, scientific exploration of the possibility that they are the product of a novel biology will have revealed that there are chemical and physical conditions in which abiotic processes mimic processes that are produced biologically on Earth; analogously, several studies showed that an excess of oxygen in the atmosphere of an exoplanet is not an infallible biosignature (e.g. \cite{Wordsworth2014,Luger2015,Narita2015}). On the other hand, if the ammonia and phosphine anomalies in Venus’ atmosphere persist in resisting efforts to provide a chemically plausible abiotic explanation then the biological explanation should be treated as potentially foreshadowing the discovery of a new biology and used as a very tentative biosignature in the search for extraterrestrial life. Detecting anomalous amounts of ammonia and phosphine in the atmosphere of yet a different planet would provide another set of (what amount to) experimental conditions for exploring the conjecture that these chemical possibilities are biotic as opposed to abiotic. 

%\funding{Please add: ``This research received no external funding'' or ``This research was funded by NAME OF FUNDER grant number XXX.'' and  and ``The APC was funded by XXX''. Check carefully that the details given are accurate and use the standard spelling of funding agency names at \url{https://search.crossref.org/funding}, any errors may affect your future funding.}

\acknowledgments{Both authors are grateful to the Leverhulme Centre for Life in the Universe (\url{https://www.lclu.cam.ac.uk/}) which organized connections and meetings, where this idea germinated. The authors also thank David H. Grinspoon and the anonymous referee for several helpful comments that significantly improved the quality of this Opinion.}

\conflictsofinterest{The authors declare no conflict of interest.} 

%%%%%%%%%%%%%%%%%%%%%%%%%%%%%%%%%%%%%%%%%%
\begin{adjustwidth}{-\extralength}{0cm}

\reftitle{References}



\end{adjustwidth}

\begin{thebibliography}{999}

\bibitem[{Hitchcock} and {Lovelock}(1967)]{Hitchcock1967}
{Hitchcock}, D.R.; {Lovelock}, J.E.
\newblock {Life detection by atmospheric analysis}.
\newblock {\em Icarus} {\bf 1967}, {\em 7},~149--159.
\newblock {\url{https://doi.org/10.1016/0019-1035(67)90059-0}}.

\bibitem[{Catling} and {Kasting}(2017)]{Catling2017}
{Catling}, D.C.; {Kasting}, J.F.
\newblock {\em {Atmospheric Evolution on Inhabited and Lifeless Worlds}};
  Cambridge University Press: Cambridge,  2017.

\bibitem[{Krissansen-Totton} \em{et~al.}(2016){Krissansen-Totton}, {Bergsman},
  and {Catling}]{Krissansen2016}
{Krissansen-Totton}, J.; {Bergsman}, D.S.; {Catling}, D.C.
\newblock {On Detecting Biospheres from Chemical Thermodynamic Disequilibrium
  in Planetary Atmospheres}.
\newblock {\em Astrobiology} {\bf 2016}, {\em 16},~39--67,
  \href{http://xxx.lanl.gov/abs/1503.08249}{{\normalfont
  [arXiv:astro-ph.EP/1503.08249]}}.
\newblock {\url{https://doi.org/10.1089/ast.2015.1327}}.

\bibitem[{Schwieterman} \em{et~al.}(2018){Schwieterman}, {Kiang}, {Parenteau},
  {Harman}, {DasSarma}, {Fisher}, {Arney}, {Hartnett}, {Reinhard}, {Olson},
  {Meadows}, {Cockell}, {Walker}, {Grenfell}, {Hegde}, {Rugheimer}, {Hu}, and
  {Lyons}]{Schwieterman2018}
{Schwieterman}, E.W.; {Kiang}, N.Y.; {Parenteau}, M.N.; {Harman}, C.E.;
  {DasSarma}, S.; {Fisher}, T.M.; {Arney}, G.N.; {Hartnett}, H.E.; {Reinhard},
  C.T.; {Olson}, S.L.;  et~al.
\newblock {Exoplanet Biosignatures: A Review of Remotely Detectable Signs of
  Life}.
\newblock {\em Astrobiology} {\bf 2018}, {\em 18},~663--708,
  \href{http://xxx.lanl.gov/abs/1705.05791}{{\normalfont
  [arXiv:astro-ph.EP/1705.05791]}}.
\newblock {\url{https://doi.org/10.1089/ast.2017.1729}}.

\bibitem[{Yung} and {Demore}(1982)]{Yung1982}
{Yung}, Y.L.; {Demore}, W.B.
\newblock {Photochemistry of the stratosphere of Venus: Implications for
  atmospheric evolution}.
\newblock {\em Icarus} {\bf 1982}, {\em 51},~199--247.
\newblock {\url{https://doi.org/10.1016/0019-1035(82)90080-X}}.

\bibitem[{Bertaux} \em{et~al.}(1996){Bertaux}, {Widemann}, {Hauchecorne},
  {Moroz}, and {Ekonomov}]{Bertaux1996}
{Bertaux}, J.L.; {Widemann}, T.; {Hauchecorne}, A.; {Moroz}, V.I.; {Ekonomov},
  A.P.
\newblock {VEGA 1 and VEGA 2 entry probes: An investigation of local UV
  absorption (220-400 nm) in the atmosphere of Venus (SO$_{2}$, aerosols, cloud
  structure)}.
\newblock {\em Journal of Geophysical Research} {\bf 1996}, {\em
  101},~12709--12746.
\newblock {\url{https://doi.org/10.1029/96JE00466}}.

\bibitem[{Ignatiev} \em{et~al.}(1999){Ignatiev}, {Moroz}, {Zasova}, and
  {Khatuntsev}]{Ignatiev1999}
{Ignatiev}, N.i.; {Moroz}, V.i.; {Zasova}, L.V.; {Khatuntsev}, I.v.
\newblock {Water vapour in the middle atmosphere of Venus:. An improved
  treatment of the Venera 15 ir spectra}.
\newblock {\em Planetary \& Space Science} {\bf 1999}, {\em 47},~1061--1075.
\newblock {\url{https://doi.org/10.1016/S0032-0633(99)00030-6}}.

\bibitem[{de Bergh} \em{et~al.}(2006){de Bergh}, {Moroz}, {Taylor}, {Crisp},
  {B{\'e}zard}, and {Zasova}]{DeBergh2006}
{de Bergh}, C.; {Moroz}, V.I.; {Taylor}, F.W.; {Crisp}, D.; {B{\'e}zard}, B.;
  {Zasova}, L.V.
\newblock {The composition of the atmosphere of Venus below 100 km altitude: An
  overview}.
\newblock {\em Planetary \& Space Science} {\bf 2006}, {\em 54},~1389--1397.
\newblock {\url{https://doi.org/10.1016/j.pss.2006.04.020}}.

\bibitem[{Oyama} \em{et~al.}(1980){Oyama}, {Carle}, {Woeller}, {Pollack},
  {Reynolds}, and {Craig}]{Oyama1980}
{Oyama}, V.I.; {Carle}, G.C.; {Woeller}, F.; {Pollack}, J.B.; {Reynolds}, R.T.;
  {Craig}, R.A.
\newblock {Pioneer Venus gas chromatography of the lower atmosphere of Venus}.
\newblock {\em Journal of Geophysical Research} {\bf 1980}, {\em
  85},~7891--7902.
\newblock {\url{https://doi.org/10.1029/JA085iA13p07891}}.

\bibitem[{Donahue} and {Hodges}(1993)]{Donahue1993}
{Donahue}, T.M.; {Hodges}, R.R.
\newblock {Venus methane and water}.
\newblock {\em Geophysical Review Letters} {\bf 1993}, {\em 20},~591--594.
\newblock {\url{https://doi.org/10.1029/93GL00513}}.

\bibitem[{Krasnopolsky}(2006)]{Krasnopolsky2006}
{Krasnopolsky}, V.A.
\newblock {Chemical composition of Venus atmosphere and clouds: Some unsolved
  problems}.
\newblock {\em Planetary \& Space Science} {\bf 2006}, {\em 54},~1352--1359.
\newblock {\url{https://doi.org/10.1016/j.pss.2006.04.019}}.

\bibitem[{Yung} \em{et~al.}(2009){Yung}, {Liang}, {Jiang}, {Shia}, {Lee},
  {B{\'e}zard}, and {Marcq}]{Yung2009}
{Yung}, Y.L.; {Liang}, M.C.; {Jiang}, X.; {Shia}, R.L.; {Lee}, C.;
  {B{\'e}zard}, B.; {Marcq}, E.
\newblock {Evidence for carbonyl sulfide (OCS) conversion to CO in the lower
  atmosphere of Venus}.
\newblock {\em Journal of Geophysical Research (Planets)} {\bf 2009}, {\em
  114},~E00B34.
\newblock {\url{https://doi.org/10.1029/2008JE003094}}.

\bibitem[{Bierson} and {Zhang}(2020)]{Bierson2020}
{Bierson}, C.J.; {Zhang}, X.
\newblock {Chemical Cycling in the Venusian Atmosphere: A Full Photochemical
  Model From the Surface to 110 km}.
\newblock {\em Journal of Geophysical Research (Planets)} {\bf 2020}, {\em
  125},~e06159.
\newblock {\url{https://doi.org/10.1029/2019JE006159}}.

\bibitem[Andreychikov \em{et~al.}(1987)Andreychikov, Akhmetshin, Korchuganov,
  Mukhin, Ogorodnikov, Petryanov, and Skitovich]{Andreychikov1987}
Andreychikov, B.; Akhmetshin, I.; Korchuganov, B.; Mukhin, L.; Ogorodnikov, B.;
  Petryanov, I.; Skitovich, V.
\newblock X-ray radiometric analysis of the cloud aerosol of Venus by the Vega
  1 and 2 probes.
\newblock {\em Cosmic Res} {\bf 1987}, {\em 25},~16.

\bibitem[{Krasnopolsky}(1989)]{Krasnopolsky1989}
{Krasnopolsky}, V.A.
\newblock {Vega mission results and chemical composition of Venusian clouds}.
\newblock {\em Icarus} {\bf 1989}, {\em 80},~202--210.
\newblock {\url{https://doi.org/10.1016/0019-1035(89)90168-1}}.

\bibitem[{Rimmer} \em{et~al.}(2021){Rimmer}, {Jordan}, {Constantinou},
  {Woitke}, {Shorttle}, {Hobbs}, and {Paschodimas}]{Rimmer2021}
{Rimmer}, P.B.; {Jordan}, S.; {Constantinou}, T.; {Woitke}, P.; {Shorttle}, O.;
  {Hobbs}, R.; {Paschodimas}, A.
\newblock {Hydroxide Salts in the Clouds of Venus: Their Effect on the Sulfur
  Cycle and Cloud Droplet pH}.
\newblock {\em The Planetary Science Journal} {\bf 2021}, {\em 2},~133,
  \href{http://xxx.lanl.gov/abs/2101.08582}{{\normalfont
  [arXiv:astro-ph.EP/2101.08582]}}.
\newblock {\url{https://doi.org/10.3847/PSJ/ac0156}}.

\bibitem[{Petkowski} \em{et~al.}(2022){Petkowski}, {Seager}, H., {Bains},
  {Ranjan}, {Rimmer}, {Buchanan}, {Agrawal}, {Mogul}, and
  {Carr}]{Petkowski2022}
{Petkowski}, J.J.; {Seager}, S.; H., G.D.; {Bains}, W.; {Ranjan}, S.; {Rimmer},
  P.B.; {Buchanan}, W.P.; {Agrawal}, R.; {Mogul}, R.; {Carr}, C.
\newblock {Venus’ Atmosphere Anomalies as Motivation for Astrobiology
  Missions}.
\newblock {\em Astrobiology} {\bf 2022}, p. submitted.

\bibitem[{Limaye} \em{et~al.}(2021){Limaye}, {Zelenyi}, and
  {Zasova}]{Limaye2021}
{Limaye}, S.S.; {Zelenyi}, L.; {Zasova}, L.
\newblock {Introducing the Venus Collection{\textemdash}Papers from the First
  Workshop on Habitability of the Cloud Layer}.
\newblock {\em Astrobiology} {\bf 2021}, {\em 21},~1157--1162.
\newblock {\url{https://doi.org/10.1089/ast.2021.0142}}.

\bibitem[{Surkov} \em{et~al.}(1973){Surkov}, {Andrejchikov}, and
  {Kalinkina}]{Surkov1973}
{Surkov}, Y.A.; {Andrejchikov}, B.M.; {Kalinkina}, O.M.
\newblock {On the content of ammonia in the Venus atmosphere based on data
  obtained from Venera 8 automatic station.}
\newblock {\em Akademiia Nauk SSSR Doklady} {\bf 1973}, {\em 213},~296--298.

\bibitem[{Surkov} \em{et~al.}(1977){Surkov}, {Andreichikov}, and
  {Kalinkina}]{Surkov1977}
{Surkov}, I.A.; {Andreichikov}, B.M.; {Kalinkina}, O.M.
\newblock {Gas-analysis equipment for the automatic interplanetary stations
  Venera-4, 5, 6 and 8.}
\newblock {\em Space Science Instrumentation} {\bf 1977}, {\em 3},~301--310.

\bibitem[{Goettel} and {Lewis}(1974)]{Goettel1974}
{Goettel}, K.A.; {Lewis}, J.S.
\newblock {Ammonia in the atmosphere of Venus.}
\newblock {\em Journal of Atmospheric Sciences} {\bf 1974}, {\em 31},~828--830.
\newblock
  {\url{https://doi.org/10.1175/1520-0469(1974)031<0828:AITAOV>2.0.CO;2}}.

\bibitem[{Mogul} \em{et~al.}(2021){Mogul}, {Limaye}, {Way}, and
  {Cordova}]{Mogul2021}
{Mogul}, R.; {Limaye}, S.S.; {Way}, M.J.; {Cordova}, J.A.
\newblock {Venus' Mass Spectra Show Signs of Disequilibria in the Middle
  Clouds}.
\newblock {\em Geophysical Research Letters} {\bf 2021}, {\em 48},~e91327,
  \href{http://xxx.lanl.gov/abs/2009.12758}{{\normalfont
  [arXiv:astro-ph.EP/2009.12758]}}.
\newblock {\url{https://doi.org/10.1029/2020GL091327}}.

\bibitem[{Bains} \em{et~al.}(2021){Bains}, {Petkowski}, {Rimmer}, and
  {Seager}]{Bains2021}
{Bains}, W.; {Petkowski}, J.J.; {Rimmer}, P.B.; {Seager}, S.
\newblock {Production of ammonia makes Venusian clouds habitable and explains
  observed cloud-level chemical anomalies}.
\newblock {\em Proceedings of the National Academy of Science} {\bf 2021}, {\em
  118},~e2110889118,  \href{http://xxx.lanl.gov/abs/2112.10850}{{\normalfont
  [arXiv:astro-ph.EP/2112.10850]}}.
\newblock {\url{https://doi.org/10.1073/pnas.2110889118}}.

\bibitem[Grinspoon(1997)]{Grinspoon1997}
Grinspoon, D.H.
\newblock {\em Venus revealed: a new look below the clouds of our mysterious
  twin planet}; Addison-Wesley: Boston,  1997.

\bibitem[{Schulze-Makuch} and {Irwin}(2002)]{Schulze-Makuch2002}
{Schulze-Makuch}, D.; {Irwin}, L.N.
\newblock {Reassessing the Possibility of Life on Venus: Proposal for an
  Astrobiology Mission}.
\newblock {\em Astrobiology} {\bf 2002}, {\em 2},~197--202.
\newblock {\url{https://doi.org/10.1089/15311070260192264}}.

\bibitem[{Schulze-Makuch} \em{et~al.}(2004){Schulze-Makuch}, {Grinspoon},
  {Abbas}, {Irwin}, and {Bullock}]{Schulze-Makuch2004}
{Schulze-Makuch}, D.; {Grinspoon}, D.H.; {Abbas}, O.; {Irwin}, L.N.; {Bullock},
  M.A.
\newblock {A Sulfur-Based Survival Strategy for Putative Phototrophic Life in
  the Venusian Atmosphere}.
\newblock {\em Astrobiology} {\bf 2004}, {\em 4},~11--18.
\newblock {\url{https://doi.org/10.1089/153110704773600203}}.

\bibitem[{Grinspoon} and {Bullock}(2007)]{Grinspoon2007}
{Grinspoon}, D.H.; {Bullock}, M.A.
\newblock {Astrobiology and Venus exploration}.
\newblock {\em Washington DC American Geophysical Union Geophysical Monograph
  Series} {\bf 2007}, {\em 176},~191--206.
\newblock {\url{https://doi.org/10.1029/176GM12}}.

\bibitem[{Greaves} \em{et~al.}(2021){Greaves}, {Richards}, {Bains}, {Rimmer},
  {Sagawa}, {Clements}, {Seager}, {Petkowski}, {Sousa-Silva}, {Ranjan},
  {Drabek-Maunder}, {Fraser}, {Cartwright}, {Mueller-Wodarg}, {Zhan},
  {Friberg}, {Coulson}, {Lee}, and {Hoge}]{Greaves2021}
{Greaves}, J.S.; {Richards}, A.M.S.; {Bains}, W.; {Rimmer}, P.B.; {Sagawa}, H.;
  {Clements}, D.L.; {Seager}, S.; {Petkowski}, J.J.; {Sousa-Silva}, C.;
  {Ranjan}, S.;  et~al.
\newblock {Phosphine gas in the cloud decks of Venus}.
\newblock {\em Nature Astronomy} {\bf 2021}, {\em 5},~655--664,
  \href{http://xxx.lanl.gov/abs/2009.06593}{{\normalfont
  [arXiv:astro-ph.EP/2009.06593]}}.
\newblock {\url{https://doi.org/10.1038/s41550-020-1174-4}}.

\bibitem[{Sousa-Silva} \em{et~al.}(2020){Sousa-Silva}, {Seager}, {Ranjan},
  {Petkowski}, {Zhan}, {Hu}, and {Bains}]{Sousa2020}
{Sousa-Silva}, C.; {Seager}, S.; {Ranjan}, S.; {Petkowski}, J.J.; {Zhan}, Z.;
  {Hu}, R.; {Bains}, W.
\newblock {Phosphine as a Biosignature Gas in Exoplanet Atmospheres}.
\newblock {\em Astrobiology} {\bf 2020}, {\em 20},~235--268,
  \href{http://xxx.lanl.gov/abs/1910.05224}{{\normalfont
  [arXiv:astro-ph.EP/1910.05224]}}.
\newblock {\url{https://doi.org/10.1089/ast.2018.1954}}.

\bibitem[{Bains} \em{et~al.}(2021){Bains}, {Petkowski}, {Seager}, {Ranjan},
  {Sousa-Silva}, {Rimmer}, {Zhan}, {Greaves}, and {Richards}]{Bains2021b}
{Bains}, W.; {Petkowski}, J.J.; {Seager}, S.; {Ranjan}, S.; {Sousa-Silva}, C.;
  {Rimmer}, P.B.; {Zhan}, Z.; {Greaves}, J.S.; {Richards}, A.M.S.
\newblock {Phosphine on Venus Cannot Be Explained by Conventional Processes}.
\newblock {\em Astrobiology} {\bf 2021}, {\em 21},~1277--1304,
  \href{http://xxx.lanl.gov/abs/2009.06499}{{\normalfont
  [arXiv:astro-ph.EP/2009.06499]}}.
\newblock {\url{https://doi.org/10.1089/ast.2020.2352}}.

\bibitem[{Snellen} \em{et~al.}(2020){Snellen}, {Guzman-Ramirez}, {Hogerheijde},
  {Hygate}, and {van der Tak}]{Snellen2020}
{Snellen}, I.A.G.; {Guzman-Ramirez}, L.; {Hogerheijde}, M.R.; {Hygate}, A.P.S.;
  {van der Tak}, F.F.S.
\newblock {Re-analysis of the 267 GHz ALMA observations of Venus. No
  statistically significant detection of phosphine}.
\newblock {\em Astronomy \& Astrophysics} {\bf 2020}, {\em 644},~L2,
  \href{http://xxx.lanl.gov/abs/2010.09761}{{\normalfont
  [arXiv:astro-ph.EP/2010.09761]}}.
\newblock {\url{https://doi.org/10.1051/0004-6361/202039717}}.

\bibitem[{Akins} \em{et~al.}(2021){Akins}, {Lincowski}, {Meadows}, and
  {Steffes}]{Akins2021}
{Akins}, A.B.; {Lincowski}, A.P.; {Meadows}, V.S.; {Steffes}, P.G.
\newblock {Complications in the ALMA Detection of Phosphine at Venus}.
\newblock {\em The Astrophysical Journal Letters} {\bf 2021}, {\em 907},~L27,
  \href{http://xxx.lanl.gov/abs/2101.09831}{{\normalfont
  [arXiv:astro-ph.EP/2101.09831]}}.
\newblock {\url{https://doi.org/10.3847/2041-8213/abd56a}}.

\bibitem[Villanueva \em{et~al.}(2021)Villanueva, Cordiner, Irwin, de~Pater,
  Butler, Gurwell, Milam, Nixon, Luszcz-Cook, Wilson, et~al.]{Villanueva2021}
Villanueva, G.; Cordiner, M.; Irwin, P.; de~Pater, I.; Butler, B.; Gurwell, M.;
  Milam, S.; Nixon, C.; Luszcz-Cook, S.; Wilson, C.;  et~al.
\newblock No evidence of phosphine in the atmosphere of Venus from independent
  analyses.
\newblock {\em Nature Astronomy} {\bf 2021}, {\em 5},~631--635.

\bibitem[{Greaves} \em{et~al.}(2021{\natexlab{a}}){Greaves}, {Richards},
  {Bains}, {Rimmer}, {Clements}, {Seager}, {Petkowski}, {Sousa-Silva},
  {Ranjan}, and {Fraser}]{Greaves2021b}
{Greaves}, J.S.; {Richards}, A.M.S.; {Bains}, W.; {Rimmer}, P.B.; {Clements},
  D.L.; {Seager}, S.; {Petkowski}, J.J.; {Sousa-Silva}, C.; {Ranjan}, S.;
  {Fraser}, H.J.
\newblock {Reply to: No evidence of phosphine in the atmosphere of Venus from
  independent analyses}.
\newblock {\em Nature Astronomy} {\bf 2021}, {\em 5},~636--639.
\newblock {\url{https://doi.org/10.1038/s41550-021-01424-x}}.

\bibitem[{Greaves} \em{et~al.}(2021{\natexlab{b}}){Greaves}, {Richards},
  {Bains}, {Rimmer}, {Sagawa}, {Clements}, {Seager}, {Petkowski},
  {Sousa-Silva}, {Ranjan}, {Drabek-Maunder}, {Fraser}, {Cartwright},
  {Mueller-Wodarg}, {Zhan}, {Friberg}, {Coulson}, {Lee}, and
  {Hoge}]{Greaves2021c}
{Greaves}, J.S.; {Richards}, A.M.S.; {Bains}, W.; {Rimmer}, P.B.; {Sagawa}, H.;
  {Clements}, D.L.; {Seager}, S.; {Petkowski}, J.J.; {Sousa-Silva}, C.;
  {Ranjan}, S.;  et~al.
\newblock {Addendum: Phosphine gas in the cloud deck of Venus}.
\newblock {\em Nature Astronomy} {\bf 2021}, {\em 5},~726--728.
\newblock {\url{https://doi.org/10.1038/s41550-021-01423-y}}.

\bibitem[{Lincowski} \em{et~al.}(2021){Lincowski}, {Meadows}, {Crisp}, {Akins},
  {Schwieterman}, {Arney}, {Wong}, {Steffes}, {Parenteau}, and
  {Domagal-Goldman}]{Lincowski2021}
{Lincowski}, A.P.; {Meadows}, V.S.; {Crisp}, D.; {Akins}, A.B.; {Schwieterman},
  E.W.; {Arney}, G.N.; {Wong}, M.L.; {Steffes}, P.G.; {Parenteau}, M.N.;
  {Domagal-Goldman}, S.
\newblock {Claimed Detection of PH$_{3}$ in the Clouds of Venus Is Consistent
  with Mesospheric SO$_{2}$}.
\newblock {\em The Astrophysical Journal Letters} {\bf 2021}, {\em 908},~L44,
  \href{http://xxx.lanl.gov/abs/2101.09837}{{\normalfont
  [arXiv:astro-ph.EP/2101.09837]}}.
\newblock {\url{https://doi.org/10.3847/2041-8213/abde47}}.

\bibitem[{Thompson}(2021)]{Thompson2021}
{Thompson}, M.A.
\newblock {The statistical reliability of 267-GHz JCMT observations of Venus:
  no significant evidence for phosphine absorption}.
\newblock {\em Monthly Notices of the Royal Astronomical Society} {\bf 2021},
  {\em 501},~L18--L22,  \href{http://xxx.lanl.gov/abs/2010.15188}{{\normalfont
  [arXiv:astro-ph.EP/2010.15188]}}.
\newblock {\url{https://doi.org/10.1093/mnrasl/slaa187}}.

\bibitem[{Greaves} \em{et~al.}(2022){Greaves}, {Rimmer}, {Richards},
  {Petkowski}, {Bains}, {Ranjan}, {Seager}, {Clements}, {Silva}, and
  {Fraser}]{Greaves2022}
{Greaves}, J.S.; {Rimmer}, P.B.; {Richards}, A.M.S.; {Petkowski}, J.J.;
  {Bains}, W.; {Ranjan}, S.; {Seager}, S.; {Clements}, D.L.; {Silva}, C.S.;
  {Fraser}, H.J.
\newblock {Low levels of sulphur dioxide contamination of Venusian phosphine
  spectra}.
\newblock {\em Monthly Notices of the Royal Astronomical Society} {\bf 2022},
  {\em 514},~2994--3001,
  \href{http://xxx.lanl.gov/abs/2108.08393}{{\normalfont
  [arXiv:astro-ph.EP/2108.08393]}}.
\newblock {\url{https://doi.org/10.1093/mnras/stac1438}}.

\bibitem[Cordiner \em{et~al.}()Cordiner, Villanueva, Wiesemeyer, Milam,
  de~Pater, Moullet, Aladro, Nixon, Thelen, Charnley, et~al.]{Cordiner2022}
Cordiner, M.; Villanueva, G.; Wiesemeyer, H.; Milam, S.; de~Pater, I.; Moullet,
  A.; Aladro, R.; Nixon, C.; Thelen, A.; Charnley, S.;  et~al.
\newblock Phosphine in the Venusian Atmosphere: A Strict Upper Limit from SOFIA
  GREAT Observations.
\newblock {\em Geophysical Research Letters}, p. e2022GL101055.

\bibitem[{Encrenaz} \em{et~al.}(2020){Encrenaz}, {Greathouse}, {Marcq},
  {Widemann}, {B{\'e}zard}, {Fouchet}, {Giles}, {Sagawa}, {Greaves}, and
  {Sousa-Silva}]{Encrenaz2020}
{Encrenaz}, T.; {Greathouse}, T.K.; {Marcq}, E.; {Widemann}, T.; {B{\'e}zard},
  B.; {Fouchet}, T.; {Giles}, R.; {Sagawa}, H.; {Greaves}, J.; {Sousa-Silva},
  C.
\newblock {A stringent upper limit of the PH$_{3}$ abundance at the cloud top
  of Venus}.
\newblock {\em Astronomy \& Astrophysics} {\bf 2020}, {\em 643},~L5,
  \href{http://xxx.lanl.gov/abs/2010.07817}{{\normalfont
  [arXiv:astro-ph.EP/2010.07817]}}.
\newblock {\url{https://doi.org/10.1051/0004-6361/202039559}}.

\bibitem[{Trompet} \em{et~al.}(2021){Trompet}, {Robert}, {Mahieux}, {Schmidt},
  {Erwin}, and {Vandaele}]{Trompet2021}
{Trompet}, L.; {Robert}, S.; {Mahieux}, A.; {Schmidt}, F.; {Erwin}, J.;
  {Vandaele}, A.C.
\newblock {Phosphine in Venus' atmosphere: Detection attempts and upper limits
  above the cloud top assessed from the SOIR/VEx spectra}.
\newblock {\em Astronomy \& Astrophysics} {\bf 2021}, {\em 645},~L4.
\newblock {\url{https://doi.org/10.1051/0004-6361/202039932}}.

\bibitem[{Omran} \em{et~al.}(2021){Omran}, {Oze}, {Jackson}, {Mehta}, {Barge},
  {Bada}, and {Pasek}]{Omran2021}
{Omran}, A.; {Oze}, C.; {Jackson}, B.; {Mehta}, C.; {Barge}, L.M.; {Bada}, J.;
  {Pasek}, M.A.
\newblock {Phosphine Generation Pathways on Rocky Planets}.
\newblock {\em Astrobiology} {\bf 2021}, {\em 21},~1264--1276.
\newblock {\url{https://doi.org/10.1089/ast.2021.0034}}.

\bibitem[{Truong} and {Lunine}(2021)]{Truong2021}
{Truong}, N.; {Lunine}, J.I.
\newblock {Volcanically extruded phosphides as an abiotic source of Venusian
  phosphine}.
\newblock {\em Proceedings of the National Academy of Science} {\bf 2021}, {\em
  118},~e2021689118.
\newblock {\url{https://doi.org/10.1073/pnas.2021689118}}.

\bibitem[{Bains} \em{et~al.}(2022{\natexlab{a}}){Bains}, {Shorttle}, {Ranjan},
  {Rimmer}, {Petkowski}, {Greaves}, and {Seager}]{Bains2022}
{Bains}, W.; {Shorttle}, O.; {Ranjan}, S.; {Rimmer}, P.B.; {Petkowski}, J.J.;
  {Greaves}, J.S.; {Seager}, S.
\newblock {Only extraordinary volcanism can explain the presence of parts per
  billion phosphine on Venus}.
\newblock {\em Proceedings of the National Academy of Science} {\bf 2022}, {\em
  119},~e2121702119.
\newblock {\url{https://doi.org/10.1073/pnas.2121702119}}.

\bibitem[{Bains} \em{et~al.}(2022{\natexlab{b}}){Bains}, {Shorttle}, {Ranjan},
  {Rimmer}, {Petkowski}, {Greaves}, and {Seager}]{Bains2022b}
{Bains}, W.; {Shorttle}, O.; {Ranjan}, S.; {Rimmer}, P.B.; {Petkowski}, J.J.;
  {Greaves}, J.S.; {Seager}, S.
\newblock {Constraints on the Production of Phosphine by Venusian Volcanoes}.
\newblock {\em Universe} {\bf 2022}, {\em 8},~54,
  \href{http://xxx.lanl.gov/abs/2112.00140}{{\normalfont
  [arXiv:astro-ph.EP/2112.00140]}}.
\newblock {\url{https://doi.org/10.3390/universe8010054}}.

\bibitem[Jordan \em{et~al.}(2022)Jordan, Shorttle, and Rimmer]{Jordan2022}
Jordan, S.; Shorttle, O.; Rimmer, P.B.
\newblock Proposed energy-metabolisms cannot explain the atmospheric chemistry
  of Venus.
\newblock {\em Nature Communications} {\bf 2022}, {\em 13},~1--10.

\bibitem[{Garvin} \em{et~al.}(2022){Garvin}, {Getty}, {Arney}, {Johnson},
  {Kohler}, {Schwer}, {Sekerak}, {Bartels}, {Saylor}, {Elliott}, {Goodloe},
  {Garrison}, {Cottini}, {Izenberg}, {Lorenz}, {Malespin}, {Ravine}, {Webster},
  {Atkinson}, {Aslam}, {Atreya}, {Bos}, {Brinckerhoff}, {Campbell}, {Crisp},
  {Filiberto}, {Forget}, {Gilmore}, {Gorius}, {Grinspoon}, {Hofmann}, {Kane},
  {Kiefer}, {Lebonnois}, {Mahaffy}, {Pavlov}, {Trainer}, {Zahnle}, and
  {Zolotov}]{Garvin2022}
{Garvin}, J.B.; {Getty}, S.A.; {Arney}, G.N.; {Johnson}, N.M.; {Kohler}, E.;
  {Schwer}, K.O.; {Sekerak}, M.; {Bartels}, A.; {Saylor}, R.S.; {Elliott},
  V.E.;  et~al.
\newblock {Revealing the Mysteries of Venus: The DAVINCI Mission}.
\newblock {\em PSJ} {\bf 2022}, {\em 3},~117,
  \href{http://xxx.lanl.gov/abs/2206.07211}{{\normalfont
  [arXiv:astro-ph.EP/2206.07211]}}.
\newblock {\url{https://doi.org/10.3847/PSJ/ac63c2}}.

\bibitem[{Seager} \em{et~al.}(2022){Seager}, {Petkowski}, {Carr}, {Grinspoon},
  {Ehlmann}, {Saikia}, {Agrawal}, {Buchanan}, {Weber}, {French}, {Klupar},
  {Worden}, and {Baumgardner}]{Seager2022}
{Seager}, S.; {Petkowski}, J.J.; {Carr}, C.E.; {Grinspoon}, D.H.; {Ehlmann},
  B.L.; {Saikia}, S.J.; {Agrawal}, R.; {Buchanan}, W.P.; {Weber}, M.U.;
  {French}, R.;  et~al.
\newblock {Venus Life Finder Missions Motivation and Summary}.
\newblock {\em arXiv e-prints} {\bf 2022}, p. arXiv:2208.05570,
  \href{http://xxx.lanl.gov/abs/2208.05570}{{\normalfont
  [arXiv:astro-ph.IM/2208.05570]}}.

\bibitem[{French} \em{et~al.}(2022){French}, {Mandy}, {Hunter}, {Mosleh},
  {Sinclair}, {Beck}, {Seager}, {Petkowski}, {Carr}, {Grinspoon}, and
  {Baumgardner}]{French2022}
{French}, R.; {Mandy}, C.; {Hunter}, R.; {Mosleh}, E.; {Sinclair}, D.; {Beck},
  P.; {Seager}, S.; {Petkowski}, J.J.; {Carr}, C.E.; {Grinspoon}, D.H.;  et~al.
\newblock {Rocket Lab Mission to Venus}.
\newblock {\em Aerospace} {\bf 2022}, {\em 9},~445,
  \href{http://xxx.lanl.gov/abs/2208.07724}{{\normalfont
  [arXiv:astro-ph.IM/2208.07724]}}.
\newblock {\url{https://doi.org/10.3390/aerospace9080445}}.

\bibitem[{Seager}(2014)]{Seager2014}
{Seager}, S.
\newblock {The future of spectroscopic life detection on exoplanets}.
\newblock {\em Proceedings of the National Academy of Science} {\bf 2014}, {\em
  111},~12634--12640.
\newblock {\url{https://doi.org/10.1073/pnas.1304213111}}.

\bibitem[Kuhn(1962)]{Kuhn1962}
Kuhn, T.S.
\newblock {\em The Structure of Scientific Revolutions}; University of Chicago
  Press,  1962; pp. Chapters VI -- VIII.

\bibitem[Stanford and Schweikert(2006)]{Stanford2006}
Stanford, P.K.; Schweikert, W.
\newblock {\em Exceeding our grasp: Science, history, and the problem of
  unconceived alternatives}; Vol.~1, Oxford University Press,  2006.

\bibitem[Evans()]{Evans2021}
Evans, B.
\newblock March 13 marks 240 years since William Herschel discovered Uranus.
\newblock {\em Astronomy Magazine}, pp.
  \url{https://astronomy.com/news/2021/03/william--herschels--discovery--of--uranus--240--years--later}.

\bibitem[NobelPrize.org(1989)]{Nobel1989}
NobelPrize.org.
\newblock Press Release: Ribonucleic acid (RNA) – a biomolecule of many
  functions.
\newblock
  \url{https://www.nobelprize.org/prizes/chemistry/1989/press-release/},  1989.
\newblock [Online; accessed 2-November-2022].

\bibitem[Oreskes(1999)]{Oreskes1999}
Oreskes, N.
\newblock {\em The rejection of continental drift: Theory and method in
  American earth science}; Oxford University Press,  1999.

\bibitem[Burnett(2007)]{Burnett2007}
Burnett, D.G.
\newblock Trying Leviathan. In {\em Trying Leviathan}; Princeton University
  Press,  2007.

\bibitem[Romero(2012)]{Romero2012}
Romero, A.
\newblock When whales became mammals: the scientific journey of cetaceans from
  fish to mammals in the history of science.
\newblock {\em New approaches to the study of marine mammals} {\bf 2012}, pp.
  3--30.

\bibitem[{Mumma} \em{et~al.}(2009){Mumma}, {Villanueva}, {Novak}, {Hewagama},
  {Bonev}, {DiSanti}, {Mandell}, and {Smith}]{Mumma2009}
{Mumma}, M.J.; {Villanueva}, G.L.; {Novak}, R.E.; {Hewagama}, T.; {Bonev},
  B.P.; {DiSanti}, M.A.; {Mandell}, A.M.; {Smith}, M.D.
\newblock {Strong Release of Methane on Mars in Northern Summer 2003}.
\newblock {\em Science} {\bf 2009}, {\em 323},~1041.
\newblock {\url{https://doi.org/10.1126/science.1165243}}.

\bibitem[Zahnle \em{et~al.}(2011)Zahnle, Freedman, and Catling]{Zahnle2011}
Zahnle, K.; Freedman, R.S.; Catling, D.C.
\newblock Is there methane on Mars?
\newblock {\em Icarus} {\bf 2011}, {\em 212},~493--503.

\bibitem[{Gillen} \em{et~al.}(2020){Gillen}, {Rimmer}, and
  {Catling}]{Gillen2020}
{Gillen}, E.; {Rimmer}, P.B.; {Catling}, D.C.
\newblock {Statistical analysis of Curiosity data shows no evidence for a
  strong seasonal cycle of martian methane}.
\newblock {\em Icarus} {\bf 2020}, {\em 336},~113407,
  \href{http://xxx.lanl.gov/abs/1908.02041}{{\normalfont
  [arXiv:astro-ph.EP/1908.02041]}}.
\newblock {\url{https://doi.org/10.1016/j.icarus.2019.113407}}.

\bibitem[{Cleland}(2019)]{Cleland2019}
{Cleland}, C.E.
\newblock {Moving Beyond Definitions in the Search for Extraterrestrial Life}.
\newblock {\em Astrobiology} {\bf 2019}, {\em 19},~722--729.
\newblock {\url{https://doi.org/10.1089/ast.2018.1980}}.

\bibitem[Cleland(2019)]{Cleland2019b}
Cleland, C.E.
\newblock {\em The Quest for a Universal Theory of Life: Searching for Life as
  we don't know it}; Vol.~11, Cambridge University Press,  2019.

\bibitem[Grinspoon(2003)]{Grinspoon2003}
Grinspoon, D.
\newblock {\em Lonely Planets: The Natural Philosophy of Alien Life};
  Ecco/HarperCollins: New York,  2003.

\bibitem[{Catling} \em{et~al.}(2018){Catling}, {Krissansen-Totton}, {Kiang},
  {Crisp}, {Robinson}, {DasSarma}, {Rushby}, {Del Genio}, {Bains}, and
  {Domagal-Goldman}]{Catling2018}
{Catling}, D.C.; {Krissansen-Totton}, J.; {Kiang}, N.Y.; {Crisp}, D.;
  {Robinson}, T.D.; {DasSarma}, S.; {Rushby}, A.J.; {Del Genio}, A.; {Bains},
  W.; {Domagal-Goldman}, S.
\newblock {Exoplanet Biosignatures: A Framework for Their Assessment}.
\newblock {\em Astrobiology} {\bf 2018}, {\em 18},~709--738,
  \href{http://xxx.lanl.gov/abs/1705.06381}{{\normalfont
  [arXiv:astro-ph.EP/1705.06381]}}.
\newblock {\url{https://doi.org/10.1089/ast.2017.1737}}.

\bibitem[{Walker} \em{et~al.}(2018){Walker}, {Bains}, {Cronin}, {DasSarma},
  {Danielache}, {Domagal-Goldman}, {Kacar}, {Kiang}, {Lenardic}, {Reinhard},
  {Moore}, {Schwieterman}, {Shkolnik}, and {Smith}]{Walker2018}
{Walker}, S.I.; {Bains}, W.; {Cronin}, L.; {DasSarma}, S.; {Danielache}, S.;
  {Domagal-Goldman}, S.; {Kacar}, B.; {Kiang}, N.Y.; {Lenardic}, A.;
  {Reinhard}, C.T.;  et~al.
\newblock {Exoplanet Biosignatures: Future Directions}.
\newblock {\em Astrobiology} {\bf 2018}, {\em 18},~779--824,
  \href{http://xxx.lanl.gov/abs/1705.08071}{{\normalfont
  [arXiv:astro-ph.EP/1705.08071]}}.
\newblock {\url{https://doi.org/10.1089/ast.2017.1738}}.

\bibitem[{Wordsworth} and {Pierrehumbert}(2014)]{Wordsworth2014}
{Wordsworth}, R.; {Pierrehumbert}, R.
\newblock {Abiotic Oxygen-dominated Atmospheres on Terrestrial Habitable Zone
  Planets}.
\newblock {\em Astrophysical Journal Letters} {\bf 2014}, {\em 785},~L20,
  \href{http://xxx.lanl.gov/abs/1403.2713}{{\normalfont
  [arXiv:astro-ph.EP/1403.2713]}}.
\newblock {\url{https://doi.org/10.1088/2041-8205/785/2/L20}}.

\bibitem[Luger and Barnes(2015)]{Luger2015}
Luger, R.; Barnes, R.
\newblock Extreme water loss and abiotic O2 buildup on planets throughout the
  habitable zones of M dwarfs.
\newblock {\em Astrobiology} {\bf 2015}, {\em 15},~119--143.

\bibitem[Narita \em{et~al.}(2015)Narita, Enomoto, Masaoka, and
  Kusakabe]{Narita2015}
Narita, N.; Enomoto, T.; Masaoka, S.; Kusakabe, N.
\newblock Titania may produce abiotic oxygen atmospheres on habitable
  exoplanets.
\newblock {\em Scientific reports} {\bf 2015}, {\em 5},~1--6.

\end{thebibliography}
\end{document}